\documentclass{pasj00}

\begin{document}
\Received{2004/12/03}
\Accepted{2005/02/28}

\SetRunningHead{T.~Tanaka et al.}{Spectral Evolution of a Luminous X-Ray 
Source (NGC~253 X21)}

\title{Spectral Evolution of a Luminous Compact X-Ray Source in NGC~253 with Chandra and XMM-Newton
\thanks{Partly based on the observations obtained with XMM-Newton, an ESA science mission with instruments and contributions directly funded by ESA Member States and NASA, USA.}}
\author{
   Takaaki \textsc{Tanaka},\altaffilmark{1,2}
   Masahiko \textsc{Sugiho},\altaffilmark{2}
   Aya \textsc{Kubota},\altaffilmark{3}
   Kazuo \textsc{Makishima},\altaffilmark{2,3} 
   \\ and
   Tadayuki \textsc{Takahashi}\altaffilmark{1,2}}
 \altaffiltext{1}{Institute of Space and Astronautical Science, JAXA, 3-1-1 
       Yoshinodai, Sagamihara, Kanagawa 229-8510}
       \email{ttanaka@astro.isas.jaxa.jp}
  \altaffiltext{2}{Department of Physics, The University of Tokyo, 7-3-1 Hongo, Bunkyo-ku, Tokyo 113-0033}
  \altaffiltext{3}{RIKEN, 2-1 Hirosawa, Wako, Saitama 351-0198}


\maketitle

\begin{abstract}
Spectral studies of a luminous X-ray source, NGC~253 X21, are presented 
based on archival Chandra and XMM-Newton data. 
The Chandra observation on 1999 December 16 detected the source at a 
bolometric luminosity of $0.3 \times 10^{39}~{\rm erg~s^{-1}}$ (assuming isotropic 
emission), while an XMM-Newton observation on 2000 June 3 revealed a short-term 
source variation in the range of (0.6--1.3)$\times 10^{39}~{\rm erg~s^{-1}}$. 
All spectra from these observations were successfully modeled by 
emission from an optically thick accretion disk. 
The  average inner disk radius was kept constant at 
$63\cdot(\cos~60^\circ/\cos~i)^{1/2}$~km, where $i$ is the disk inclination, 
and did not vary significantly, 
while the disk inner temperature changed in the range of 0.9--1.4~keV.
Assuming that this object is an accreting Schwarzschild 
black hole, and that the disk inner 
radius coincides with its last stable orbit, 
the mass of the black hole is estimated to be $\sim 7~M_\odot$. The disk 
luminosity corresponds to (30--120)$\cdot(\cos~60^\circ/\cos~i)$~\% 
of the Eddington limit of this black hole. 
Therefore, this luminous X-ray source, NGC~253 X21, is understood 
consistently to be an accreting stellar mass black hole in the standard disk state. 
\end{abstract}

\section{Introduction}
Since the era of the Einstein Observatory, many compact X-ray sources
have been detected in nearby galaxies (e.g., \cite{Fab89}).
Sources with X-ray luminosities below $\sim 2\times10^{38}~{\rm erg~s^{-1}}$, 
which corresponds to the Eddington limit 
for a neutron star of $1.4~M_\odot$, can be accreting neutron stars,
while those with higher luminosities can be considered 
as black hole binaries (hereafter, BHB).
The most luminous class of compact objects, 
known as ultra-luminous compact X-ray sources (ULXs; \cite{Max00}), 
show X-ray luminosity of $\sim10^{39}$--$10^{40}~\rm{erg~s^{-1}}$. 
Supposing that ULXs obey the Eddington limit,
their high luminosities imply that they are accreting black holes
as massive as 10--100$\, M_{\odot}$ or more.
Such luminous X-ray objects, which are rather rare in our Galaxy and LMC, 
are suggested to reside preferentially 
in galaxies with high star-formation activities \citep{GGS04}.
Therefore, starburst galaxies provide a good opportunity to study stellar black holes
of various masses. 

In studying Galactic/Magellanic BHBs, 
intensity-correlated spectral changes have been used as a key diagnostic tool. 
When the luminosity is above $\sim 3\%$ 
of the Eddington limit, a BHB is often in the standard-disk 
(alternatively called high or soft) 
state, wherein the X-ray spectrum is successfully described by a so-called multi-color 
disk model (MCD model; \cite{Mi84}), predicted by the theory of optically thick 
standard accretion disks \citep{S&S}. The bolometric disk luminosity, $L_{\rm disk}$, 
varies as $\propto {T_{\rm in}}^4$, where $T_{\rm in}$ is the highest color 
temperature of the disk. 
Equivalently, the calculated innermost disk radius, $R_{\rm in}$, stays constant, 
at a value consistent with the radius of the last stable orbit around the black hole. 
As the source becomes more luminous (close to the Eddington limit), $R_{\rm in}$ 
apparently becomes variable, as $R_{\rm in} \propto {T_{\rm in}}^{-1}$, 
probably because the disk makes a transition into a slim disk (\cite{KME01}; \cite{KM04}). 
A very similar behavior has been observed for several of the most luminous ULXs, which 
show MCD-like spectra (\cite{Miz01}; \cite{Sug03}; \cite{KDM02}). 
It is therefore of keen interest to look for such intensity-correlated spectral changes 
among extragalactic BHB candidates, in particular including those with lower 
luminosities, and to examine whether the behavior indicates standard disks or 
slim-disks.

With this perspective, we analyzed the XMM-Newton and Chandra
data of NGC~253 X21 (\cite{Vol99}; also named Source~5 in \cite{FT84}).
The nearby Sc-type edge-on galaxy NGC~253 
(at a distance of 2.58~Mpc; \cite{PC88}) is
one of the well-known starburst galaxies with prominent diffuse X-ray emission 
(e.g., \cite{St02}). Many point-like X-ray sources have also been detected since 
the era of the Einstein Observatory. 
Among them, X21 is one of the brightest sources, located
approximately $4^{\prime}$ southwest from the center of this galaxy.
\citet{Vol99} showed that this source is 
time-variable by analyzing ROSAT data, and 
\citet{Pi01} analyzed the XMM-Newton data of X21 
and reported that its luminosity varied by a factor of $\sim 2$ during 
a 40~ks observation. 
In a large sample of ULXs observed with Chandra and XMM-Newton 
\citep{Sug03}, this source showed a large luminosity variation in the range of 
$\sim10^{38}~\rm{erg~s^{-1}}$--$10^{39}~\rm{erg~s^{-1}}$. 
Therefore, this object deserves a detailed study in 
comparison with Galactic/Magellanic BHBs.

\begin{figure}
 \begin{center}
  \FigureFile(85mm,70mm){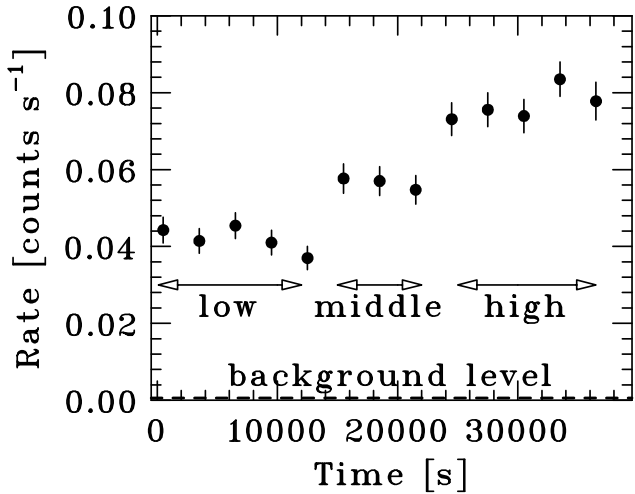}
\end{center}
\caption{The 0.5--10~keV light curve of X21 (in counts s$^{-1}$), 
averaged over the two MOS cameras of XMM-Newton. 
The background is not subtracted.}
\label{fig:lc_mos}
\end{figure}

\section{Observations and Data Reduction}
The XMM-Newton \citep{Jan01} observation of NGC~253, to be utilized 
in the present paper, was carried out on 2000 June 3. 
The data, retrieved from the XMM-Newton archive, were processed using 
the Science Analysis System (SAS), version 5.4.1. 
To create response matrices, the software package rmfgen was used. 
Since X21 was located on the CCD gap of the PN detector \citep{Str01} 
during the observation, 
the present paper utilizes only the MOS (MOS~1 and MOS~2; \cite{Tur01}) data. 
For each EPIC camera, events with pattern 0 to 12 were accumulated over 
a region of $15^{\prime \prime}$, centered on the X21 image. 
A total of 35~ks of good data were obtained for the two MOS 
cameras by discarding the time region of a short background time flare. 
Figure \ref{fig:lc_mos} shows a 0.5--10~keV light curve with an 
average count rate of each MOS camera of $0.06~{\rm cts~s}^{-1}$, 
which roughly gives a 0.5--10~keV flux of $1 \times 10^{-12}~{\rm erg~s^{-1}~cm^{-2}}$. 
As shown by the dashed line at the bottom, the background counts were negligible 
($\sim 1\%$ of the source counts in this range). 
The signal count rate varied by a factor of two, as already reported by \citet{Pi01}. 

The present work also utilized a Chandra  \citep{Wei00} observation of NGC~253, 
performed on 1999 December 16. 
The Chandra data consisted of a net exposure of 14~ks, when the 
target source was on the S3 chip. The data reduction was carried out using 
CIAO software version 2.3, and the response matrix was made with mkrmf. 
The source events were extracted from a circular region 
of $3^{\prime \prime}.7$ radius around the source centroid. 
The average 0.5--10~keV count rate 
was 0.06~${\rm cts~s}^{-1}$, which corresponded to a 0.5--10~keV energy flux of 
$3 \times 10^{-13}~{\rm erg~s^{-1}~cm^{-2}}$, a factor of three 
lower than the average flux observed with XMM-Newton. 
There was no significant time variation beyond the Poisson noise. 

\begin{figure}
 \begin{center}
 \FigureFile(85mm,70mm){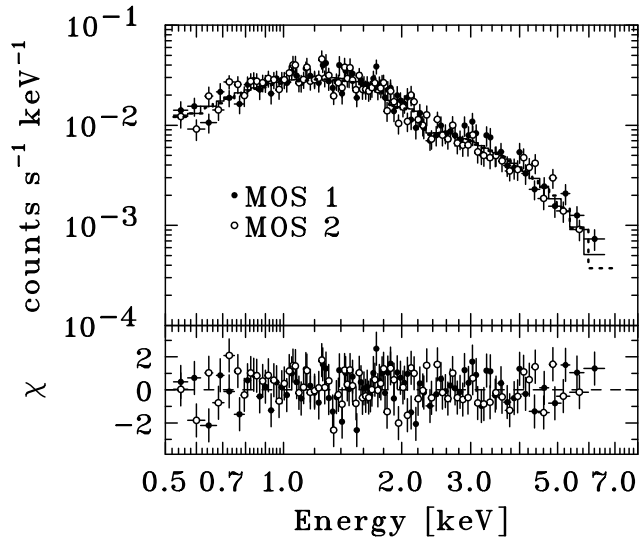}
\end{center}
\caption{Time-averaged XMM-Newton spectra 
of X21, together with the data residuals from the best-fit MCD model.}
\label{fig:spec_mos_av}
\end{figure}

\begin{table*}
\caption{The best-fit parameters of the energy spectra of NGC~253 X21\footnotemark[$*$] }
\label{tab:fit_par}
\begin{center}
\begin{tabular}{lllccc}
\hline
\hline
Observations & Model & Parameter & $N_{\rm H}$\footnotemark[$\dagger$] & $f_x$\footnotemark[$\ddagger$] & $\chi^2/\nu$\\
\hline
XMM-Newton, average & MCD & $T_{\mathrm{in}}$\footnotemark[$\S$]$=1.28^{+0.07}_{-0.06},~R_{\mathrm{in}}$\footnotemark[$\|$]$=59\pm6$ & $6\pm2$ & 9.38 & 149.9/152  \\
 & Power law & $\Gamma$\footnotemark[\#]$= 2.01^{+0.09}_{-0.08}  $ & $27\pm4$ & 10.7 & 175.9/152  \\

XMM-Newton, high-flux period & MCD & $T_{\mathrm{in}}  = 1.4\pm0.2,~R_{\mathrm{in}}  = 60^{+16}_{-12}$ & $11\pm7$ & 14.0 & 27.5/42  \\

XMM-Newton, middle-flux period & MCD & $T_{\mathrm{in}}  = 1.3\pm0.2,~R_{\mathrm{in}}  = 63^{+21}_{-16}$ & $6\pm6$ & 9.47 & 31.2/33  \\

XMM-Newton, low-flux period& MCD & $T_{\mathrm{in}}  = 1.2^{+0.1}_{-0.2},~R_{\mathrm{in}}  = 60^{+13}_{-10}$ & $1 ^{+5}_{-1}$ & 6.37 & 46.7/49  \\

Chandra, average & MCD & $T_{\mathrm{in}}  = 0.9\pm0.1,~R_{\mathrm{in}}  = 69^{+21}_{-14}$ & $2^{+4}_{-2}$ & 3.15 & 23.8/27  \\
 & Power law & $ \Gamma  = 2.2^{+0.3}_{-0.2} $ & $20^{+7}_{-6}$ & 3.94 & 25.4/27  \\
\hline
\multicolumn{6}{@{}l@{}}{\hbox to 0pt{\parbox{180mm}{\footnotesize
      \par\noindent
      \footnotemark[$*$] Errors represent 90\% confidence.
     \par\noindent
      \footnotemark[$\dagger$] Hydrogen column density in the unit of $10^{20}~\mathrm{cm}^{-2}$
      \par\noindent
      \footnotemark[$\ddagger$] 0.5--10~keV X-ray flux in the unit of $10^{-13}~\mathrm{erg}~\mathrm{s}^{-1} \, \mathrm{cm}^{-2}$.
      \par\noindent
      \footnotemark[$\S$] In the unit of keV.
      \par\noindent
      \footnotemark[$\|$] In the unit of km.
      \par\noindent
      \footnotemark[\#] Photon index.
     }\hss}}
\end{tabular}
\end{center}
\end{table*}

\section{Analyses and Results}
\subsection{Averaged Spectra}
Figure \ref{fig:spec_mos_av} shows the time-averaged spectra of X21 
obtained with the XMM-Newton MOS cameras. 
The background events 
were extracted from a source-free circular region of $2^{\prime}$ radius 
on the same chip as the source. 
Following the previous studies on ULXs 
(e.g. \cite{Max00}), the MCD model and a single power law (PL) model were 
fit to the spectra. 
As summarized in table \ref{tab:fit_par}, the MCD 
model successfully describes the observed spectra with 
$\chi^2/\nu = 149.9/152$, 
while the PL model is less successful with $\chi^2/\nu = 175.9/152$. 
The value of the hydrogen column density, $N_{\rm H}$, obtained with the 
MCD fit, is roughly consistent with the line-of-sight Galactic value of 
$N_{\rm H} = 1.4 \times 10^{20}~{\rm cm^{-2}}$ \citep{DL90}, 
while that with the PL model is by a factor 
of 20 higher than the Galactic value. 
Furthermore, the value of $N_{\rm H}$ estimated from the MCD fit 
is consistent with the value from a spectral analysis of diffuse emission 
in this region of NGC~253 reported by \citet{St00}. 
Thus, the MCD model explains the data 
more appropriately than the PL model. 
Employing a correction factor for the inner-boundary condition of 
$\xi = 0.41$ \citep{K98} and a color-to-effective temperature 
correction factor of $\kappa \sim 1.7$ \citep{S&T}, 
the best-fit MCD model yields the innermost disk radius, 
$R_{\rm in} = (59\pm6)\cdot \zeta^{1/2}$~km 
with $\zeta\equiv \cos~60^\circ/\cos~i$, 
where $i$ is the inclination of the system. 
By using a formula for the disk bolometric luminosity, $L_{\rm disk}$, 
by \citet{Max00},  
\begin{eqnarray}
L_{\rm disk} = 4 \pi \left( \frac{R_{\rm in}}{\xi} \right)^2 \sigma \left( \frac{T_{\rm in}}{\kappa} \right)^4,
\end{eqnarray}
where $\sigma$ is the Stefan--Boltzmann constant, 
$L_{\rm disk}$\ is then calculated as  
$8.6 \times 10^{38}\cdot \zeta~\rm{erg~s^{-1}}$, which 
corresponds to the Eddington limit of a $5.8~M_\odot$ object. 
An MCD plus PL model with 
a fixed photon index of $\Gamma = 2.0$ gives essentially the same result 
within 90\% errors, because the PL component in this two-component fit 
is not significant. 

\begin{figure}
 \begin{center}
   \FigureFile(85mm,70mm){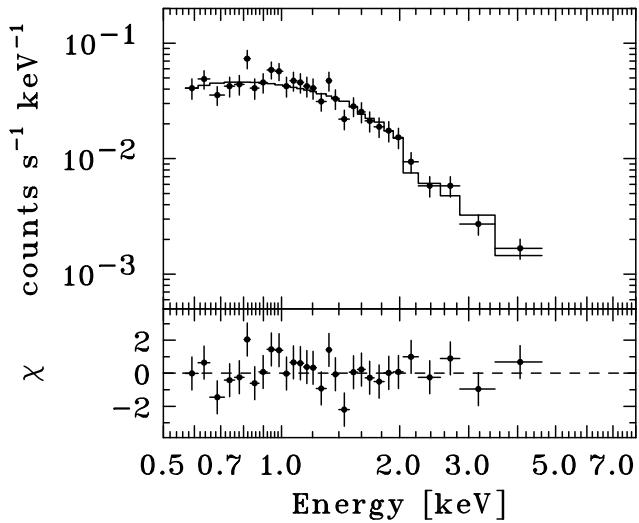}
\end{center}
\caption{Chandra ACIS spectrum of X21, together with the best-fit 
MCD model. The bottom panel shows the fit residuals.}
\label{fig:spec_acis}
\end{figure}

The Chandra ACIS spectrum of X21 was analyzed in the same manner. 
Figure~\ref{fig:spec_acis} shows the obtained source spectrum, 
after subtracting the background spectrum, which was extracted from 
a source-free circular region of $30^{\prime \prime}$ radius on the same 
chip as the source. 
The spectrum was fit with the MCD model 
and the PL model, as was done for the XMM-Newton data. 
As shown in table~\ref{tab:fit_par}, 
the spectrum can be described by either model because of the poor statistics.  
However, the MCD fit is again more reasonable from the viewpoint of $N_{\rm H}$. 
The Chandra data thus yield 
$R_{\rm in} = (69^{+21}_{-14})\cdot \zeta^{1/2}$~km
and $L_{\rm disk} = 2.9 \times 10^{38}\cdot \zeta~\rm{erg~s^{-1}}$. 
Between the XMM-Newton and the Chandra observations, 
the value of $R_{\rm in}$ was kept almost constant at $\sim 60 \cdot \zeta^{1/2}$ km, 
while $L_{\rm disk}$ changed by a factor of three. 

\begin{figure}
 \begin{center}
   \FigureFile(85mm,70mm){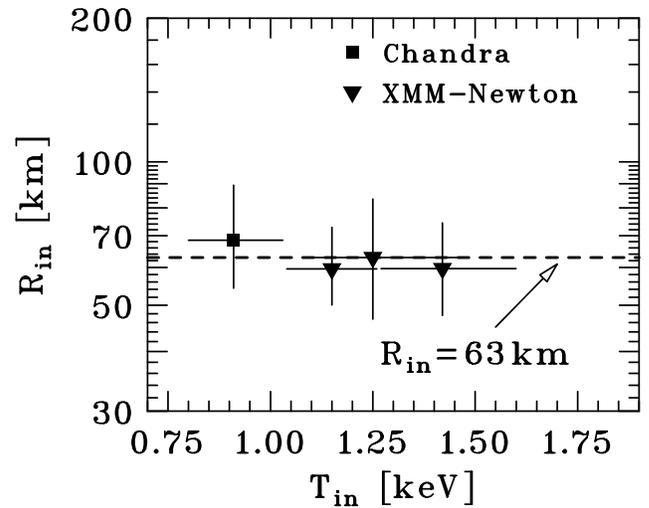}
\end{center}
\caption{Relation between the highest color temperature of the disk ($T_{\rm in}$) and 
the calculated innermost disk radius ($R_{\rm in}$) of NGC~253 X21. 
The dashed line represents $R_{\rm in} = 63~{\rm km}$. A disk inclination of 
$i=60^\circ$ is assumed.}
\label{fig:Tin_Rin}
\end{figure}

\subsection{Intensity-Correlated Change of the Spectrum}
As shown in figure~\ref{fig:lc_mos} and reported by \citet{Pi01}, 
X21 exhibited a significant intensity
variation by a factor of two during the XMM-Newton observation.
Accordingly, the entire observational span was split into three periods; covering 
low-flux, middle-flux, and high-flux periods, as denoted by the three arrows in the figure.
Spectra constructed from each period were investigated with the absorbed  
MCD model 
to clarify the spectral evolution in response to the intensity change. 
As shown in table~\ref{tab:fit_par}, this model reproduced each spectrum very well. 

Figure~\ref{fig:Tin_Rin} shows the values of $R_{\rm in}$ 
against $T_{\rm in}$, based on the intensity-sorted XMM-Newton spectra 
 (filled inverted triangles) and the Chandra spectrum (filled square).
It clearly shows that the value of $R_{\rm in}$ remained constant during 
the significant variation of $T_{\rm in}$. 
The data points are consistent with the behavior of BHBs in the standard-disk state, 
where $R_{\rm in}$ remained constant. 
The average value of  $R_{\rm in}$ is determined as 
$(63^{+9}_{-7})\cdot \zeta^{1/2}~{\rm km}$.  

\begin{figure}
 \begin{center}
  \FigureFile(85mm,80mm){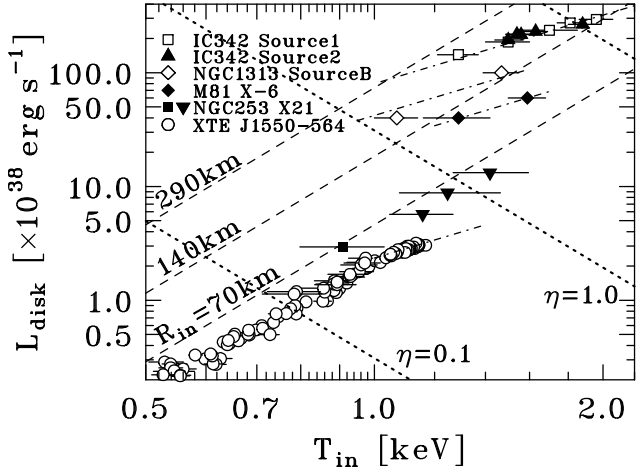}
 \end{center}
\caption{Scatter plot between the disk bolometric luminosity ($L_{\rm disk}$) 
and the highest color temperature ($T_{\rm in}$) obtained from the MCD-model fit. 
All data points of NGC~253 X21 refer to the present work. Those of other ULXs, 
IC~342 Source~1, IC~342 Source~2, M~81 X-6 and NGC~1313 Source~B, were 
obtained with ASCA \citep{Miz01}. The data points of the Galactic BHB, 
XTE~J$1550-564$, are taken from \citet{KM04}, and are 
shifted downward for clarity by a factor of two. The inclination is assumed to be 
$60^{\circ}$ for all data. The dashed lines represent 
$L_{\rm disk}\propto {T_{\rm in}}^4$, hence mean constant value of 
$R_{\rm in}$, and the dot-dashed lines represent 
$L_{\rm disk}\propto {T_{\rm in}}^2$. The parameter $\eta$ specifies 
the disk bolometric luminosity normalized to the Eddington luminosity
 ($\eta = L_{\rm disk}/L_{\rm Edd}$).
For NGC~253 X21, the same symbols with figure~\ref{fig:Tin_Rin} are used.}
\label{fig:hr}
\end{figure}

In figure~\ref{fig:hr}, the calculated values of $L_{\rm disk}$ are plotted 
against $T_{\rm in}$ for each period, in order to examine the spectral 
behavior against the luminosity. 
This diagram shows that the data points satisfy the relation 
$L_{\rm disk}\propto {T_{\rm in}}^4$ as $L_{\rm disk}$ changes over 
(0.3--1.3) $\times 10^{39}\cdot \zeta~{\rm erg~s^{-1}}$. 
In fact, a logarithmic slope of $3.6{\pm0.6}$ is obtained when these data points are fitted with a power law.

\section{Discussion}
Through the combined use of archival Chandra and XMM-Newton data, 
an overall luminosity change by a factor of three was observed from NGC~253 
X21.
The X-ray spectra at all intensity levels were successfully described by the MCD model,
and the highest disk luminosity reached
$1.3\times10^{39}\cdot \zeta~{\rm erg~s^{-1}}$.
This luminosity, on one hand, is the lowest among those of ULXs.
On the other hand, this source is shinning at a higher luminosity than most 
Galactic/Magellanic sources.
Except for a few examples, such as GRS~$1915+105$ and XTE~J$1550-564$ 
in its outburst peak, 
no compact X-ray source in our Galaxy has so far exhibited a luminosity
higher than $1\times10^{39}~{\rm erg~s^{-1}}$ (e.g., \cite{G&D}; \cite{Zycki}).

In order to study the spectral behavior of X21 compared 
with luminous ULXs and XTE~J$1550-564$, which is a representative Galactic BHB, 
their $T_{\rm in}$--$L_{\rm disk}$ relations are plotted together in
figure~\ref{fig:hr}. 
While the luminosities of luminous ULXs are observed to vary as 
$\propto {T_{\rm in}}^2$,
that of NGC~253 X21 varied as $L_{\rm disk} \propto {T_{\rm in}}^4$. 
The latter relation is considered to be a signature of standard accretion disks 
\citep{S&S}, as has been observed from a fair number of BHBs, including 
XTE~J$1550-564$, LMC~X-3 \citep{Eb93}, GS~$2000+25$ \citep{Taki91}, 
and GS~$1124-684$ \citep{Og92}. 
The present results thus provide one of the first confirmations of the standard-disk 
property in BHB candidates outside the local group of galaxies. 

As shown in Section 3, the value of $R_{\rm in}$ of X21 was estimated to be 
$63\cdot \zeta^{1/2}$~km.
The innermost stable orbit of a non-rotating BH is given as $3R_{\rm s}$, where 
$R_{\rm s}=2GM/c^2$ is the Schwarzschild radius. 
Thus, the mass of the central black hole of this source is estimated to 
be $7\cdot \zeta^{1/2}~M_\odot$.
This is a very common value found among BHBs 
in our Galaxy and LMC (e.g., \cite{M&R05}).
Thus, the observed disk luminosity of 
(0.3--1.3)$ \times10^{39}\cdot \zeta~{\rm erg~s^{-1}}$ corresonds to
(30--$120) \cdot \zeta$~\% of the Eddington limit for the inferred BH mass. 
Therefore, NGC~253 X21 can be understood consistently as an accreting BH with 
an ordinary stellar mass, in which a standard accretion disk is radiating at a fair fraction 
of the Eddington limit. In particular, the observed values of 
$T_{\rm in} = 0.9$--1.4~keV are reasonable for the inferred BH mass and luminosity, 
thus making X21 free from the problem of ``too high a disk temperature'' 
observed from the most luminous ULXs \citep{Max00}. 

According to the present results, the least-luminous class of the entire ULXs 
population may contain ordinary BHBs. Similar examples include 
M~33 X-8 (\cite{Taka94}; \cite{Max00}), NGC~2403 X-3 \citep{Kot00}, 
and NGC~4449 Source~2 \citep{Miy04}. 
Future studies of luminosity-correlated spectral evolution with a larger 
sample of extragalactic X-ray point sources will elucidate the detailed 
composition of ULXs. 
\\
\\
The authors would like to thank Chris Done for her valuable comments. 
T.~Tanaka is supported by research fellowships of the Japan Society for the Promotion 
of Science for Young Scientists.

\end{document}